# Data Processing For Atomic Resolution EELS

Paul Cueva,[*] Robert Hovden,[*] Julia A. Mundy,[*] Huolin L. Xin,[**] and David A. Muller[*,***]

[*] School of Applied and Engineering Physics, Cornell University, Ithaca, NY 14853
[**] Department of Physics, Cornell University, Ithaca, NY 14853
[***] Kavli Institute at Cornell for Nanoscale Science, Ithaca, NY 14853

CORRESPONDING AUTHOR:

Robert Hovden
212 Clark Hall, Cornell University, Ithaca, NY 14853
Tel: 607 255-0654
Fax: 607 255-7658
E-mail: rmh244@cornell.edu

**ABSTRACT**

The high beam current and sub-angstrom resolution of aberration-corrected scanning transmission electron microscopes has enabled electron energy loss spectroscopic (EELS) mapping with atomic resolution. These spectral maps are often dose-limited and spatially oversampled, leading to low counts/channel and are thus highly sensitive to errors in background estimation. However, by taking advantage of redundancy in the dataset map one can improve background estimation and increase chemical sensitivity. We consider two such approaches—linear combination of power laws and local background averaging—that reduce background error and improve signal extraction. Principal components analysis (PCA) can also be used to analyze spectrum images, but the poor peak-to-background ratio in EELS can lead to serious artifacts if raw EELS data is PCA filtered. We identify common artifacts and discuss alternative approaches. These algorithms are implemented within the Cornell Spectrum Imager, an open source software package for spectroscopic analysis.

**INTRODUCTION**

Elemental mapping via spectral analysis has proven extremely useful, especially in the field of scanning transmission electron microscopy (STEM). Techniques such as energy dispersive x-ray (EDX), cathodoluminescence, and electron energy loss spectroscopy (EELS) all provide spatially resolved spectroscopic information. Spectral imaging, in which many spectra are acquired as the electron probe is rastered across the specimen, forms a 2D spectral map—stored in a datacube of information (Hunt & Williams, 1991; Jeanguillaume & Colliex, 1989). The number of scanned points and signal-to-noise ratio (SNR) of a spectral image is greatly limited by the amount of signal, the instrument stability, and the user's time. Fortunately, a new generation of aberration corrected electron microscopes has not only provided sub-angstrom lateral resolutions, but the large probe forming apertures (Batson, Dellby, & Krivanek, 2002; Bosman, *et al.*, 2007; Okunishi, *et al.*, 2006) and improved collection optics (Krivanek et al., 2008) also permit higher beam currents for rapid acquisition of spectroscopic signals down to atomic resolution (Botton, Lazar, & Dwyer, 2010; D. A Muller, 2009; D A Muller, L Fitting Kourkoutis, *et al.*, 2008). This has, in part, fueled the growing popularity and the acquisition of spectral images.

It is now possible to acquire thousands of spectra in a minute. As a result, spectral data is typically oversampled spatially. This is most clearly demonstrated with atomic resolution EELS mapping, where pixel dimensions are often ten times smaller than the probe size to provide a smooth and visually pleasing image, as well as reducing sensitivity to scan and high tension instabilities. Even at the nanometer scale, changes in background behavior can be slowly varying; resulting in a spatially oversampled background signal. Approaches to exploiting the redundancy can range from simple smoothing and filtering, to multivariate statistical analysis (Trebbia & Bonnet, 1990) and filtering, both of the background (Bosman, *et al.*, 2007) and the spectra (Bosman, *et al.* 2006). This redundancy of information also offers opportunities for better background modeling—improving the size (in pixels), detection limits (in elemental concentration), and quality (in terms of SNR) of a spectral image (D. R. Liu & Brown, 1987). We find that improved background estimation can help reduce noise and artifacts not only for the obvious background extrapolations and interpolative models, but also for multivariate analysis. Here we consider two simple approaches for improved background estimation at low count

rates—linear combination of power laws (LCPL) and local background averaging (LBA) that can be combined with standard approaches to data processing including simple background extrapolation and signal integration, interpolative modeling of the background and edge combined (Verbeeck & Van Aert, 2004), and multivariate methods such as principal components analysis – PCA (Bosman, *et al.*, 2006; Trebbia & Bonnet, 1990) and multivariate curve resolution – MCR (Haaland, *et al.*, 2009).

Proper analysis of spectral data requires software for extracting the chemical signatures present in a spectrum. However, the commercial software available for spectrum analysis remains expensive, complicated and often not transparent to the internal workings and approximations made. Worse, licensing terms generally restrict the software to a single computer tied to the data-acquisition microscope or a handful of machines. For user facilities, educational institutes, or any other setting where multiple users on a single tool can be expected, the limited availability of software becomes the bottleneck to data analysis, user training, and throughput. Faced with several hundred users logging over 10,000 hours on our instruments at Cornell, we developed a universal data analysis tool that could be freely distributed, would run on all computers, and in order to minimize training, would present the same user interface for imaging, EELS, and EDX data analysis. Our goal was software that could be run with minimal instruction or training—this meant ensuring all options needed for basic analysis were immediately visible to the user, and also limiting those options to avoid overwhelming a new user. This software package, the Cornell Spectrum Imager (CSI), freely downloadable from http://code.google.com/p/cornell-spectrum-imager/ provides an open source platform for intuitive, advanced approaches to data analysis and processing of spectral images. It includes the background estimation and PCA algorithms discussed in this paper, as well as a graphical interface for principal components analysis.

## METHODS & RESULTS

### *Extracting EELS Core-loss Edge*

Electron energy loss spectroscopy (EELS), in which inelastically scattered electrons are sorted and detected according to their energies, provides detailed information about the chemical species, bonding, and structure of materials down to the atomic scale. Most often, the information of interest is contained in the core-loss energy edges. These edges appear in the EELS spectrum with a shape and energy onset uniquely defined by a specimen's excitation of core level electrons to the available density of states in the conduction band, and modified by the electron-hole interaction. The edges lie on a background signal that can be produced by contributions due to valence excitations, core level excitations, plural scattering events, and combinations thereof. Despite the complexity, it has been shown that the background often follows an inverse power law (Leapman, 2004) (R.F. Egerton, 2011), as seen in Figure 1.

The signal is most often obtained after the background has been modeled and subtracted over the edge of interest. The subtracted signal can then be integrated over a specified window (Joy & Maher, 1981). Alternatives to signal integration have also been demonstrated, most often in the form of component analysis. Component analysis, such as principle component analysis (PCA), independent component analaysis (ICA), or linear discriminant analysis (LDA), separates every spectrum in the dataset into additive subcomponents (Friedman, 1989)(Pearson, 1901). The components can be used to characterize, or smooth, information in the core-loss region. Since the components are determined from the entire dataset, they take advantage of redundancy in the data and may provide better SNR than simple integration. However, many of the unbiased multivariate data analysis methods such as the popular principle component analysis (PCA) (Bonnet, 1999; Bosman, *et al.*, 2009) can fail to detect local behavior, such as interface states, under typical conditions, although they do lead to dramatic reductions in the apparent noise for the surrounding bulk material.

PCA is an optimal algorithm in the sense that it provides the least number of orthogonal vectors needed to capture a given percentage of the image variance. While PCA performs very well on x-ray spectral maps and secondary ion mass spectroscopy where the peak to background ratio is very high, it can produce serious but hard to recognize artifacts for EELS. The poor peak-to-background ratio in EELS means that much of that variance arises from changes in slope and shape of the background, rather than the edge of interest. The result is that the usual approach of using a scree plot to determine the number of significant components can grossly underestimate the number of significant components needed to describe the data. Figure 2 shows the consequences of PCA filtering raw data to remove noise—here retaining 10 components, 7 more than indicated necessary by the scree plot. The true structure of the Al-K edge is lost and replaced by an artifact that tracks the tails of the preceding La-M edge instead (Fig. 2). For EDS and SIMS where backgrounds are very low compared to the signal, weighting the data by two-way scaling can improve the situation (Keenan & Kotula, 2004; Haaland, Jones, *et al.*, 2009).

Unfortunately for EELS, weighting often fails where most needed: in the limit of a vanishingly small signal on a large background, the scaling asymptotes back to the unweighted approximation. This effect is particularly pronounced when examining PCA filtered EELS fine structure (Fig. 3), which is often systematically distorted, with peaks shifting as much an electron volt. We have also found cases where the interface states themselves have been filtered away completely (Fig. 3c). The failure is not of PCA itself, but rather the global metric of total image variance. For instance, in an *NxN* pixel image, a line of interface states only accounts for 1/*N* of the total image pixels. As $N \rightarrow \infty$, the fraction of image variance contained in the interface states tends to 0, and its PCA rank will drop below that of bulk noise components. Employing *a priori* knowledge of model-based approaches (Verbeeck & Van Aert, 2004), or local metrics such as spatially resolved residuals often overcome these problems. Local minimizations, such as Multivariate Curve Resolution (Haaland, Jones, *et al.*, 2009) can also be effective. Reducing changes in the EELS edge from background variation can be accomplished by subtracting a background fitted to the pre-edge region. This improves PCA performance by emphasizing only variance in principle components relating to the core-loss edge structure. Similar arguments may be made for ICA or LDA, suggesting that accurate background subtraction is important to many approaches to EELS signal analysis. In general, we would strongly recommend against PCA filtering raw EELS data, but rather to subtract the background before attempting any such analysis, and to keep in mind that any PCA filtered result is likely to biased against detecting interfacial changes, especially if they are small.

### *Least Squares Estimation*

As a result of the relatively high background level, the accuracy of the background extrapolation directly—and significantly—affects the error in the signal of interest. Therefore careful EELS analysis requires accurate characterization of the pre-edge signal and faithful background extrapolation over the EELS edge of interest. Traditionally, background extrapolation is accomplished by least-squares curve fitting of the pre-edge spectrum. Although most often edges will lie on a power law background (R F Egerton, 1975), situations may arise where regions are better described by another function—e.g. linear fits over small energy window when edges overlap (Rez, 1983) or exponential fits to low energy loss regions at moderate thicknesses. However, assuming the background can be modeled as a power law, we can express it as:

$$b(E) = c\, E^{-r}$$

Improving the background estimation error, and ultimately the signal-to-noise ratio of the final extracted signal, is accomplished by reducing the variance in parameters $c$ and $r$ (R.F. Egerton, 1982). The background error will grow rapidly as the integration region becomes more distant from the fit region (Pun, et al., 1985). This limits the integration range to be near the edge onset. Therefore improving background fits has a two-fold benefit for signal extraction by reducing the error in the background estimation and also enabling larger integration windows.

The most common implementation of a power law fit is a linear least squares fit to the log-log transform of the experimental spectrum (R.F. Egerton, 2011). For point-by-point and line profile EELS maps, the counts per channel tend to be large and this approach is usually well behaved. For 2-dimensional EELS maps, the counts per channel can be very low. After dark correction of the spectra, it is not unusual to find zero or negative values. Such values are outside the domain of the log transform and therefore cannot be fitted by this simple least squares approach. Replacing such "bad" values with a positive definite constant, as is commonly done in standard software, will bias the fit, making it extremely unstable. In our own work on EELS maps of nanoparticles on thin and porous supports, this instability forced us to consider alternative approaches.

*Linear Combination of Power Laws*

For a given estimator, it is possible to improve the estimate of the background by exploiting prior knowledge of the data set at hand. For instance, knowing that physical power law exponents, $-r_i$, should be negative (decaying). One intuitive approach is to use a linear combination of power laws (LCPL). In this method, all backgrounds in a spectrum image are assumed to be accurately described by a linear combination of power laws:

$$b(E) = \sum_i c_i\, E^{-r_i}$$

Where $i$ specifies the number of power laws being used, E is the energy, $r_i$ is the specified power law parameter, and $c_i$ is the scalar coefficient determined by the best fit to the background. The power law parameters, $r_i$, form the user-specified basis, and must be chosen to best model the dataset of interest. For x-ray absorption spectroscopy, Victoreen demonstrated that the background can be accurately described by two fixed power laws (Victoreen, 1943). This spectroscopy is closely related to EELS and in our experience it is often the case that backgrounds in an EELS spectrum image can be characterized by two power laws:

$$b(E) = c_1\, E^{-r_1} + c_2\, E^{-r_2}$$

Thus, once appropriate power laws have been determined for the dataset, it is a linear least squares minimization problem to determine the two scalar coefficients, $c_1$ and $c_2$. This approach has the advantage of avoiding the log transforms of the experimental data needed for the simple power law fit, and caused its failure at low count rates. While the dimensionality of the problem (2 parameters) has not been reduced from a simple power law fit of the form $cE^{-r}$, it has greatly improved the stability of the system—in effect setting upper and lower bounds to the background's power laws. The burden is upon the user to choose appropriate power law coefficients, which requires a little knowledge of the system, or dataset, at hand. For example, with x-ray absorption, a common choice for a physical basis is 3 to 4. However, because of plural scattering in the EELS signal, the coefficients are typically around 3, and we expect them to fall in the range between 0 and 4.

One may prefer to take a more generic approach to determining the power law coefficients for LCPL. This can be effectively done by choosing power laws from a histogram of the power law coefficients that have been fit to each background in the spectrum image. We found, choosing the decaying power law coefficients at the 5% and 95% extremes (or 20/80% if very noisy) of the dataset do well at describing all power law behavior in the spectrum image. In this regard, one is considering the information of the entire dataset in order to process each individual spectrum. This global *a priori* information enables the improved background extrapolation.

The linear combination of two power laws appears robust over a range of specimens and edges— requiring only that sensible power laws are chosen. Picking the 5/95% extremes of all power law parameters in a spectral data set provided results that were robust and free from noticeable artifacts. When applied to background subtracted atomic resolution EELS spectra, where the integrated core-loss edge provides a chemical concentration map, there is a small but noticeable improvement in the SNR. This can be seen in Figure 4a, for the Cu-L, Mn-L and O-K, edge maps. The improvements in SNR become more noticeable as the count rates drop—i.e. the most improvement for Cu then Mn then O.

Additionally, the LCPL behaves well at low count rates and in vacuum regions by preventing the spurious divergent exponents that result from taking the logarithm of zero or negative count channels in a simple linear least squares power law estimation. Figure 5a shows what can happen when a simple power law fit is applied to a nanoparticle suspended over vacuum. Divergent extrapolations are often fit to the noise level in the vacuum region, resulting in wild extremes in the integrated intensities. LCPL corrects this, as visible in Figure 5b, by only allowing physically sensible negative power law coefficients. The method can potentially fail if the power laws are poorly chosen, or if a large energy window is selected in a sample with many regions of widely difference thickness. It is generally a good idea (for all approaches) to examine the background-subtracted data set to ensure the pre-edge region now fluctuates about zero, and there are no systematic deviations.

### *Local Background Averaging*

Local background averaging (LBA) provides another approach to improved background modeling. Backgrounds that vary slowly with position can be averaged with those from neighboring spectra to obtain an accurate representation of the background at a given position. With this method, a background model is fit to the locally-averaged background signal at every position. The average of multiple spectra will have less Poisson noise than its constituent spectra.

The reduced noise in the averaged background signal enables a more reliable background fit and extrapolation. Because the similarity in background behavior is more likely to decrease with radial distance from a given position, we implement a Gaussian averaging that is characterized by its full width half maximum (FWHM). For a FWHM of 2 pixels, there will be approximately a two-fold improvement in the background's SNR. In fact, because the integrated average background counts grows with the square of averaging Gaussian FWHM and the background's SNR increases with square root of the counts (assuming Poisson statistics), there is a linear relationship that holds such that the LBA SNR improves approximately proportional to the FWHM of Gaussian. For even a FWHM of a few pixels, the dramatic increase in the background's SNR can have noticeable improvements in the background extrapolation and final signal. Note that this improvement in SNR is for the background estimate itself, not the edge of interest, which is still limited by the counting statistics of the spectrum. LBA will show improved results only when it is the error in background estimation that dominates the noise.

However, if the radius of averaging extends over two dramatically different backgrounds, the algorithm begins to fail. This typically limits the averaging FWHM to a few pixels. Local averaging can easily create artifacts when significant changes in the background's power law behavior occurs spatially less than the FWHM of averaging—*e.g.* abrupt interfaces, nano-particle edges, sample to vacuum transitions. However—as often is the case—when maps oversample with respect to changes in the background, local averaging can be used free of artifacts.

Atomic resolution EELS maps are often over sampled—with pixel dimensions smaller than the probes transfer limit. In this case, LBA works exceptionally well at estimating the background signal. When the core-loss edges are integrated to form a chemical map, local background averaging can provide a dramatic improvement in the image contrast and signal-to-noise. Figure 4a, shows the improvement of local averaging over simple power law fitting and LCPL in an atomic resolution dataset. The local averaging provides a clearly visible atomic resolution Cu map that was previously lost in the noise. Although, the Cu signal in Figure 4a is a particularly dramatic example of how advantageous local averaging can be for atomic resolution images, the effects are still noticeable in other atomic maps (as seen in the Mn and O maps of Figure 4). When compared to smoothing by weighted PCA with a sufficient number of components, LBA and PCA have converged to similar compositional results (Figure 7), except LBA avoids the distortions to the EELS fine structure observed in PCA.

Despite being well suited for atomic resolution images, local background averaging can be applied to systems at other scales. Figure 5c shows how LBA improves the signal of a nanoparticle EELS map by reducing noise and increasing contrast. However, unlike atomic resolution images, which often sample beyond the probes contrast transfer limit, nm-resolution images are undersampled so there may be no correlation between neighboring pixels. Rapid variations in the background can lead to artifacts in the LBA approach. One has to always be prudent when using LBA to characterize sharp interfaces at this scale. When dramatic changes in background occur over regions larger than the FWHM of the Gaussian averaging, then the LBA signal is no longer a representative background. This usually occurs in the vicinity of interfaces—often the region of interest. For a nanoparticle, this occurs at the edges of the particles. Figure 6 shows the onset of edge artifacts with the FWHM of the Gaussian average is increased up to 20 pixels (0.133 nm/pixel). By a FWHM of 9 pixels, there is a noticeable ringing around the edge of the particle—a low intensity region just off the particle and a high intensity region just on the particle. This ringing is an artifact that can arise from LBA and may lead a researcher to reach incorrect conclusions about the core shell structure of their particle. In the case of LBA on nanoparticle shell structures, one must be cautious when making claims about features smaller than the FWHM of the local averaging.

**DISCUSSION**

*Cornell Spectrum Imager*

Accompanying this paper, we present a free, open-source software tool for spectral analysis of electron energy loss (EELS), energy dispersive x-ray (EDX), or cathodoluminescence (CL) data—the Cornell Spectrum Imager (CSI) (Fig. 8). CSI can be downloaded from http://code.google.com/p/cornell-spectrum-imager/. CSI aims to provide an efficient, intuitive user interface that also includes unique features. Built as a plugin for the ImageJ platform, CSI offers an approachable graphical interface supported by standard image analysis software. Users on PC, Mac, or Linux can work with multi-gigabyte one, two, or three-dimensional data-sets. In its current realization, CSI can read DigitalMicrograph (.dm3), Emispec, image stack, or datacube files containing a single spectrum, line profile, or map. Once the data has been read into ImageJ, the spectrum analyzer allows the option of point, line, rectangle, circle, and freehand region selection on spectrum images enables analysis of irregularly shaped features. CSI includes LCPL and local averaging for background modeling. Additional features currently include weighted and unweighted principal component analyses (PCA) that allow for smoothing and higher-level analysis of fine structure variations. CSI is optimized for generating single or composite chemical maps from three-dimensional spectral data sets, either by integration or principal components.

*Alternative Background Estimation*

When using very large energy windows it is not correct to assume that the error in each channel is equal. This is especially the case for backgrounds in the first few hundred electron-Volts of a spectrum. To overcome this, iterative weighted least squares approaches can be used to incorporate the changing in variance over the background (Pun, *et al.*, 1985). Furthermore, a maximum likelihood estimation was shown to outperform all least squares estimation approaches—with the greatest benefit when counts are low (Unser, *et al.*, 1987). Egerton has also discussed a two-window approach to improving background estimation (R. Egerton, 2002). Although these techniques were not implemented in with this work, they can readily be combined with LCPL and LBA. As CSI is as an open source project, it lends itself to public implementation of these algorithms.

**CONCLUSIONS**

The detection limits and signal-to-noise ratios of images extracted from spectroscopic mapping depends highly on the signal processing methods. This is especially true for core-loss EELS, in which pre-edge power law background modeling can greatly affect the accuracy and range of the extrapolated background. We have aimed to improve EELS background characterization by two approaches—both of which utilize *a priori* knowledge: linear combination of power laws and local background averaging. LCPL works well over a range of specimens, but is particularly useful when low background counts would normally lead to wildly fluctuating backgrounds. LBA works well when the background has been spatially oversampled. The two are not mutually exclusive and can be combined to give optimal results. They also avoid the distortions to the EELS fine structure observed with PCA. These algorithms, and others, have been implemented through open-source software based on the ImageJ platform. The Cornell Spectrum Imager provides a modifiable and extendable EELS analysis toolset that is available, free, to the microscopy community.

**ACKNOWLEDGEMENTS**

We would like to give particular thanks to Gregory Jefferis for his kind contributions to the DM3 reader ImageJ plugin as well as Peter Ercius for assistance with the multidimensional DM3 file formats. We also acknowledge helpful feedback from Pinshane Huang, Lena Fitting-Kourkoutis, and Earl Kirkland (Cornell) and Marion Stevens-Kalceff, Chee Chia, Stephen Joseph, Paul Munroe (UNSW) regarding the development of the Cornell Spectrum Imager. RH was supported by the Semiconductor Research Corporation and the Center for Nanoscale Systems, an NSF NSEC (NSF #EEC-0117770, 0646547). PC was supported by DOE BES Award #DE-SCOO02334. HLX and JAM were supported by Energy Materials Center at Cornell, an Energy Frontier Research Center (DOE BES award #DE-SC0001086). JAM was also supported by a NDSEG fellowship. This work made use of the electron microscopy facility of the Cornell Center for Materials Research (CCMR) with support from the National Science Foundation Materials Research Science and Engineering Centers (MRSEC) program (DMR 1120296) and NSF IMR-0417392.

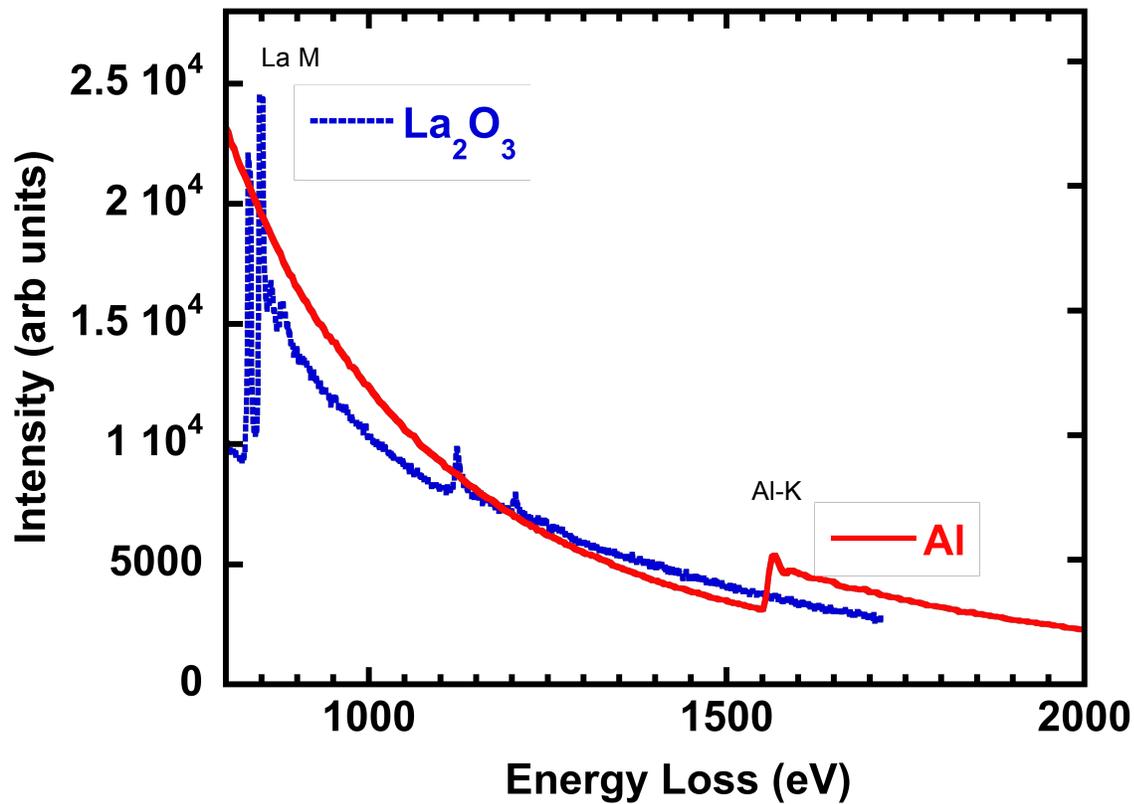

**Figure 1.** EELS spectra of the La-M and Al-K edges show the difference in power-law behavior that can occur in a specimen—the post La-M edge decay is clearly less rapid than the Al-K edge. Characterization of background behavior throughout a sample often requires multiple power-laws.

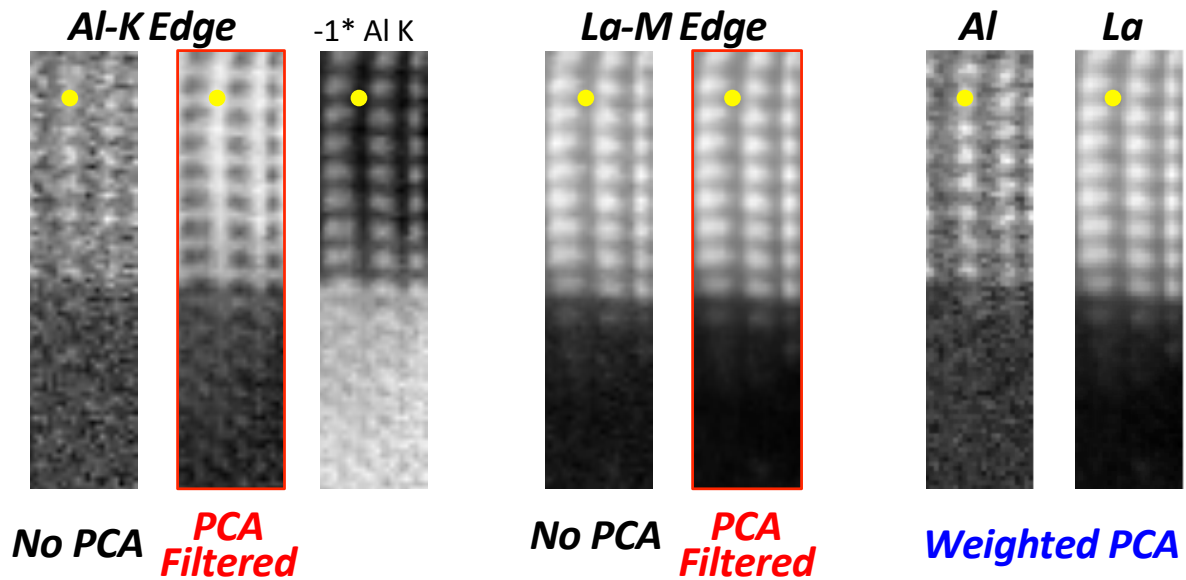

**Figure 2.** The Al-K and La-M Edges recorded across a SrTiO$_3$/LaAlO$_3$ interface. The power-law subtracted and edge integration is applied to raw data (no PCA) and PCA-filtered (1$^{st}$ ten components – 7 more than suggested by the scree plot) data sets. While the unfiltered Al-K edge shows the Al lattice correctly, the PCA-filtered data shows a contrast pattern that tracks the La-M edge contrast reversed, along with a false periodicity in the SrTiO$_3$ substrate. A weighted PCA (again 10 components) is able to restore a more plausible Al-K map. However, as shown in figure 3, the weighted PCA has introduced artifacts into the shape of the EELS fine structure and filtered out interface state.

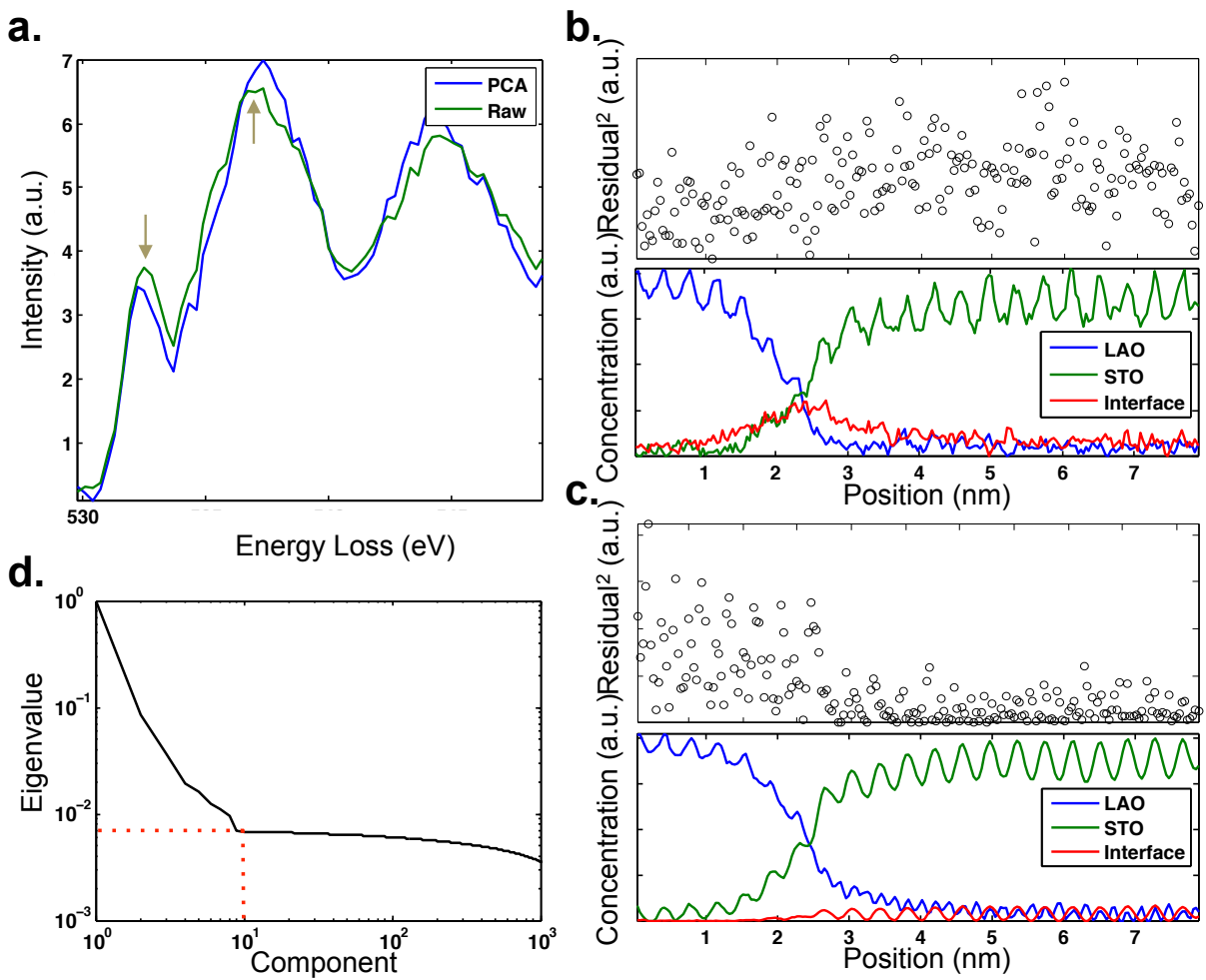

**Figure 3.** a) The O-K edge summed over 20 spectra recorded at a SrTiO$_3$/LaAlO$_3$ interface. If the raw data is 10-component PCA filtered prior to background subtraction, the shape of fine structure is altered, with the first and second peaks shifting strongly in opposite directions by -0.4 eV, +0.4 eV respectively. b,c) Three-component multivariate curve resolution (MCR) of the O-K edge across a SrTiO$_3$/LaAlO$_3$ interface. The MCR fit to the raw data (b) shows a clear interfacial component that is also apparent in the raw spectra, but is lost by weighted PCA filtering (c) the raw data prior to background subtraction. The MCR residuals appear well behaved (random) for the raw data (b) but structure remains in the filtered data (c). The weighted data (c) also displays unphysical intensity oscillations in the LAO and "interface" components. Scree plot (d) shows that almost all the variance is contained within the first 9 components (dashed line).

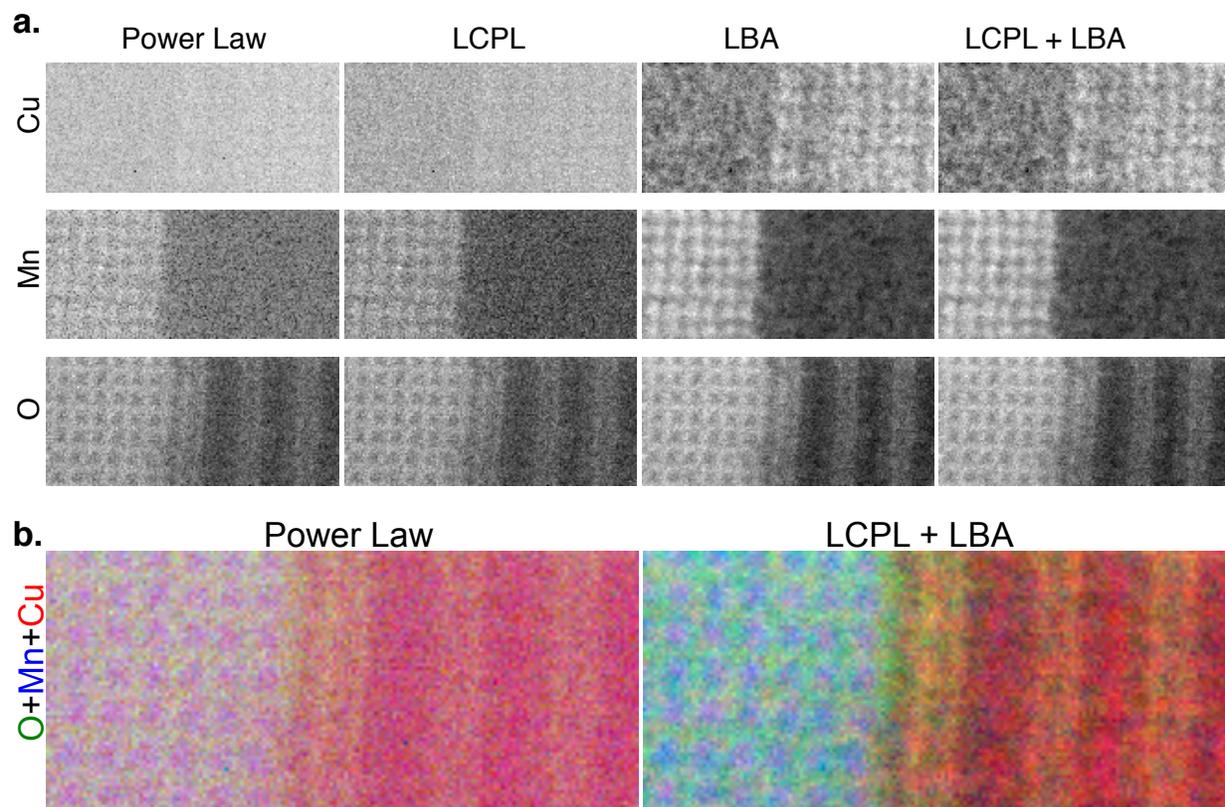

**Figure 4.** Atomic resolution chemical maps of an YBCO/Manganite interface obtained from an EELS spectroscopic image. a.) Elemental maps showing side by side comparison of different background subtraction methods b.) RGB map displaying overall improvement from using traditional power law to LCPL with LBA. Data courtesy of Lena Fitting-Kourkoutis and Jak Chakalian.

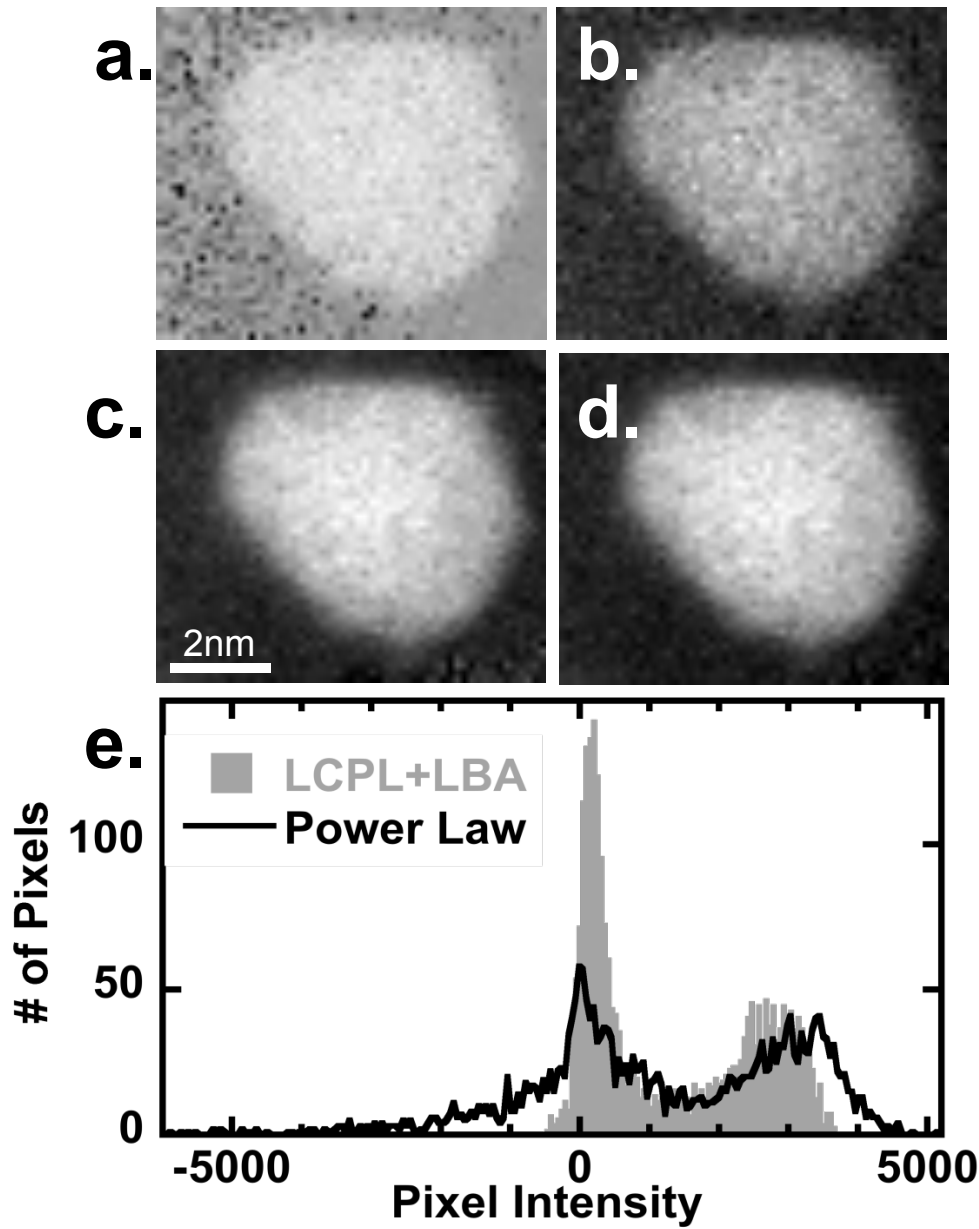

**Figure 5.** Different background subtraction algorithms on a Pt-M edge from a Pt3Co nanoparticle. The peak/background ratio is good, but the number of counts per channel is low. a) Power law fit, b) linear combination of 2 power laws (LCPL), c) 3-pixel radius local background average (LBA) for the power law exponent, d) Combining both the LCPL of (b) and the LBA of (c), e) Histograms of image intensities, showing the reduced scatter of (d) compared to the basic power law (a).

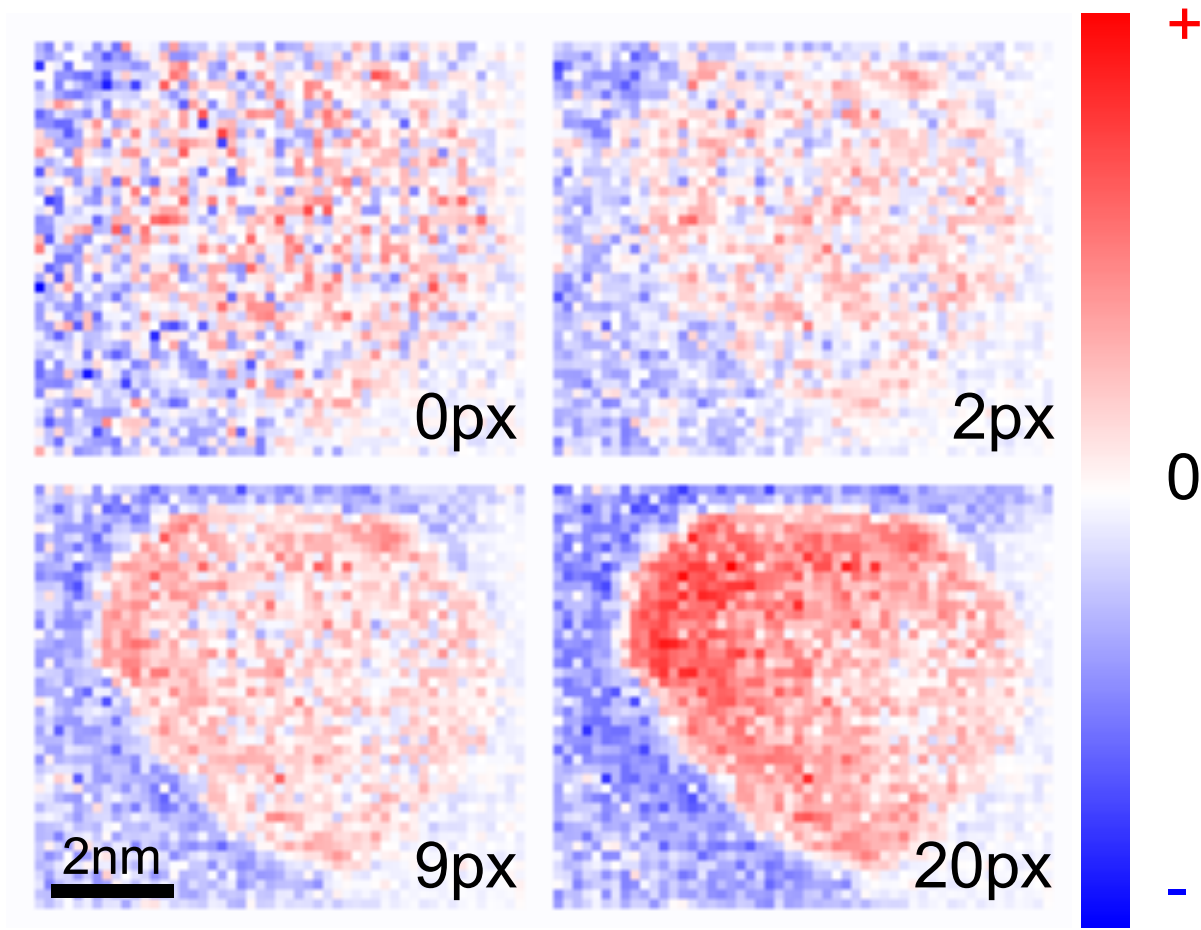

**Figure 6.** Map showing LBA integrated Pt-N edge intensity in a Pt$_3$Co particle on carbon support. The Pt-N edge has a much a larger background compared to the Pt-M edge. Increasing the size of the averaging kernel can introduce ringing – here visible as a negative (blue) ring around the particle for the 9 and 20 pixel FWHM maps. The true background should be around zero (white). Such rings could be mistaken for an unphysical core-shell structure that is not present in the Pt-M edge shown in figure 5.

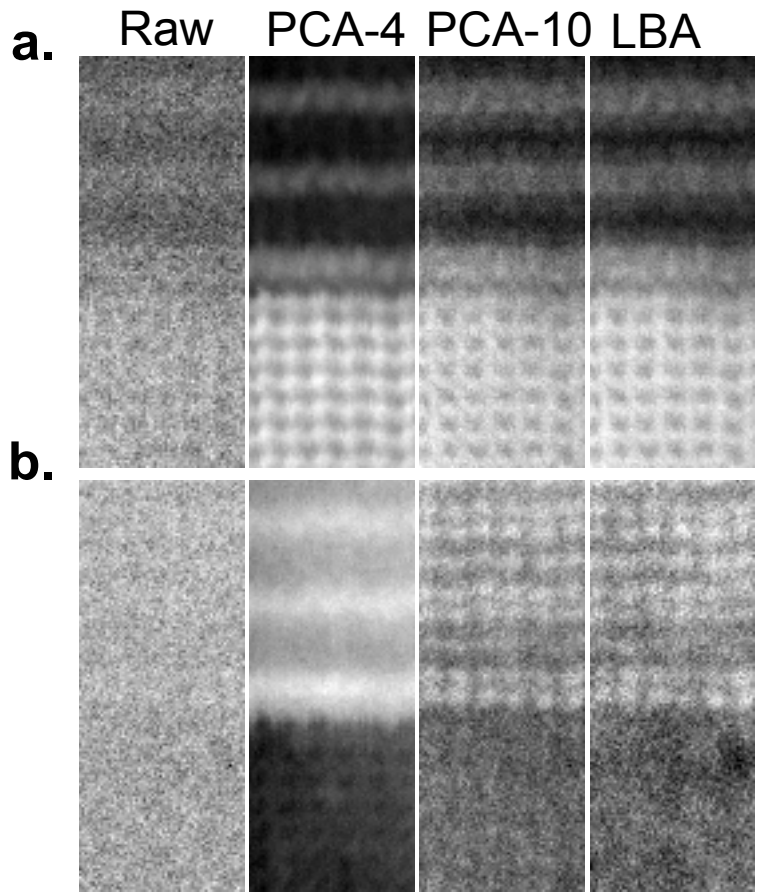

**Figure 7.** Comparison of noise reduction methods that exploit redundancy in the measured dataset for (a) the power-law-subtracted O-K and (b) Cu-L edges from the YBCO/Manganite interface(Raw) and again simple power law fits after weighted PCA filtering with 4 (PCA4) and 10 (PCA10) components, and finally the raw data after LBA and LCPL is applied instead (LBA). With 4 principal components, the maps do not reflect the physical positions of the atoms, while with 10 components (4 more than suggested by it scree plot), the PCA results comes closer to resembling the more plausible LBA/LCPL result.

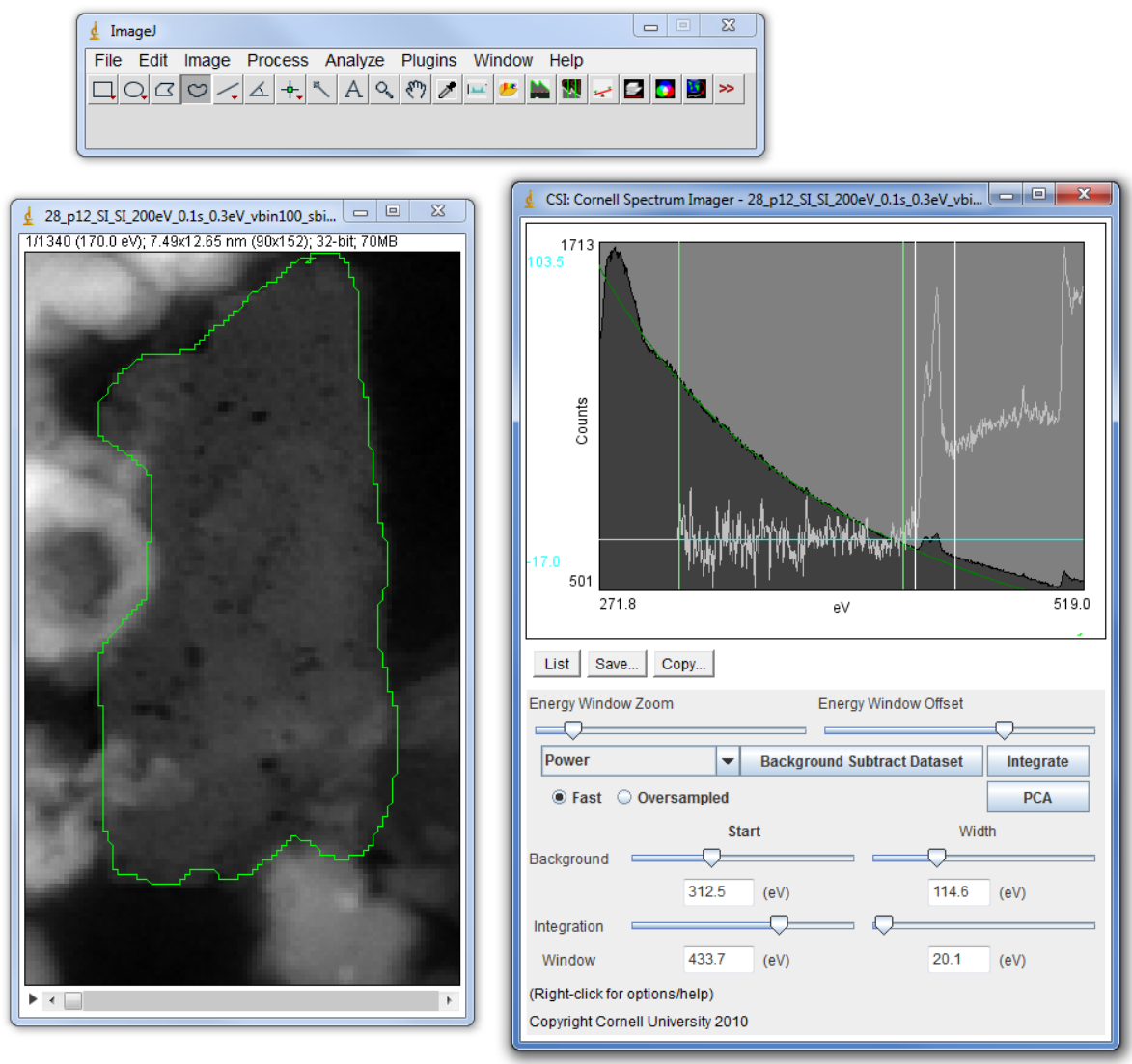

**Figure 8.** Screenshot of CSI performing analysis of a 3D EELS dataset. Averaged energy loss spectrum (*right, dark gray*) from selected freehand region of dataset (*left, green*) is shown with power-law background extrapolation and autoscaled background-subtracted spectrum. By setting appropriate background and integration windows (*right, vertical lines*), one can immediately obtain a chemical map from the integrated intensity of each spectrum contained at every pixel.

REFERENCES

Batson, P. E., Dellby, N., & Krivanek, O. L. (2002). Sub-Angstrom resolution using aberration corrected electron optics. *Nature*, *418*, 617-620.

Bonnet, N. (1999). Extracting information from sequences of spatially resolved EELS spectra using multivariate statistical analysis. *Ultramicroscopy*, *77*(3-4), 97-112. doi:10.1016/S0304-3991(99)00042-X

Bosman, M., Keast, V., Garcia-Munoz, J., Findlay, S., & Allen, L. (2007). Two-dimensional mapping of chemical information at atomic resolution. *Physical review letters*, *99*(8), 86102. APS. doi:10.1103/PhysRevLett.99.086102

Bosman, M., Watanabe, M., Alexander, D. T. L., & Keast, V. J. (2006). Mapping chemical and bonding information using multivariate analysis of electron energy-loss spectrum images. *Ultramicroscopy*, *106*(11-12), 24-32. doi:10.1016/j.ultramic.2006.04.016

Botton, G. A., Lazar, S., & Dwyer, C. (2010). Elemental mapping at the atomic scale using low accelerating voltages. *Ultramicroscopy*, *110*(8), 926-934. doi:10.1016/j.ultramic.2010.03.008

Egerton, R F. (1975). INELASTIC-SCATTERING OF 80 KEV ELECTRONS IN AMORPHOUS CARBON. *PHILOSOPHICAL MAGAZINE*, *31*(1), 199-215. doi:10.1080/14786437508229296

Egerton, R. (2002). Improved background-fitting algorithms for ionization edges in electron energy-loss spectra. *Ultramicroscopy*, *92*(2), 47-56. doi:10.1016/S0304-3991(01)00155-3

Egerton, R.F. (2011). *Electron Energy-Loss Spectroscopy in the Electron Microscope*. Boston, MA: Springer US. doi:10.1007/978-1-4419-9583-4

Egerton, R.F. (1982). A revised expression for signal/noise ratio in EELS. *Ultramicroscopy*, *9*, 387-390.

Friedman, J. (1989). Regularized Discriminant Analysis. *Journal of the American Statistical Association*, *84*(405), 165-175.

Haaland, D. M., Jones, H. D. T., Van Benthem, M. H., Sinclair, M. B., Melgaard, D. K., Stork, C. L., Pedroso, M. C., Liu, P., Brasier, A. R., Andrews, N. L., & Lidke, D. S. (2009). Hyperspectral Confocal Fluorescence Imaging: Exploring Alternative Multivariate Curve Resolution Approaches. *Applied Spectroscopy*, *63*(3), 271-279. OSA.

Hunt, J. A., & Williams, D. B. (1991). Electron energy-loss spectrum-imaging. *Ultramicroscopy*, *38*(1), 47-73. doi:10.1016/0304-3991(91)90108-I

Jeanguillaume, C., & Colliex, C. (1989). Spectrum-image: The next step in EELS digital acquisition and processing. *Ultramicroscopy*, *28*(1-4), 252-257. Elsevier. doi:10.1016/0304-3991(89)90304-5

Joy, D. C., & Maher, D. M. (1981). The quantification of electron energy loss spectra. *Journal of Microscopy*, *124*, 37-48.

Keenan, M. R., & Kotula, P. G. (2004). Accounting for Poisson noise in the multivariate analysis of ToF-SIMS spectrum images. *Surface and Interface Analysis*, *36*(3), 203-212. doi:10.1002/sia.1657

Krivanek, O. L., Corbin, G. J., Dellby, N., Elston, B. F., Keyse, R. J., Murfitt, M. F., Own, C. S., et al. (2008). An electron microscope for the aberration-corrected era. *Ultramicroscopy*, *108*(3), 179-195. doi:10.1016/j.ultramic.2007.07.010

Leapman, R. (2004). *Transmission Electron Energy Loss Spectrometry in Materials Science and The EELS Atlas*. (C. C. Ahn, Ed.). Weinheim, FRG: Wiley-VCH Verlag GmbH & Co. KGaA. doi:10.1002/3527605495

Liu, D. R., & Brown, L. M. (1987). Influence of some practical factors on background extrapolation in EELS quantification. *Journal of Microscopy*, *147*, 37-49.

Muller, D. A. (2009). Structure and bonding at the atomic scale by scanning transmission electron microscopy. *Nature materials*, *8*(4), 263-70. Nature Publishing Group. doi:10.1038/nmat2380

Muller, D A, Kourkoutis, L Fitting, Murfitt, M., Song, J. H., Hwang, H. Y., Silcox, J., Dellby, N., et al. (2008). Atomic-scale chemical imaging of composition and bonding by aberration-corrected microscopy. *Science*, *319*(5866), 1073-6. doi:10.1126/science.1148820

Okunishi, E., Sawada, H., Kondo, Y., & Kersker, M. (2006). Atomic Resolution Elemental Map of EELS with a Cs Corrected STEM. *Microscopy and Microanalysis*, *12*(S02), 1150-1151.

Pearson, K. (1901). On Lines and Planes of Closest Fit to Systems of Points in Space. *Philosophical Magazine*, *2*(6), 559-572.

Pun, T., & Ellis, J. (1985). Weighted least squares estimation of background in EELS imaging. *Journal of microscopy*, 93-100. Retrieved from http://www.ncbi.nlm.nih.gov/pubmed/3973919

Rez, P. (1983). *Detection limits and error analysis in energy loss spectrometry*. *Microbeam Analysis* (p. 153). San Francisco Press.

Trebbia, P., & Bonnet, N. (1990). EELS elemental mapping with unconventional methods, I. Theoretical basis: Image analysis with multivariate statistics and entropy concepts. *Ultramicroscopy*, *34*, 165-178.

Unser, M., Ellis, J., & Pun, T. (1987). Optimal background estimation in EELS. *Journal of microscopy*. Retrieved from http://www.ncbi.nlm.nih.gov/pubmed/3585991

Verbeeck, J., & Van Aert, S. (2004). Model based quantification of EELS spectra. *Ultramicroscopy*, *101*(2-4), 207-24. doi:10.1016/j.ultramic.2004.06.004

Victoreen, J. A. (1943). Probable X-Ray Mass Absorption Coefficients for Wave-Lengths Shorter Than the K Critical Absorption Wave-Length. *Journal of Applied Physics*, *14*(2), 95. doi:10.1063/1.1714956